# Emittance Minimization for Aberration Correction II: Physics-informed Bayesian Optimization of an Electron Microscope*

Desheng Ma*

*School of Applied and Engineering Physics,*

*Cornell University, Ithaca, NY 14853, USA*

Steven E. Zeltmann

*Platform for the Accelerated Realization, Analysis, and Discovery of Interface Materials and*

*School of Applied and Engineering Physics,*

*Cornell University, Ithaca, NY 14853, USA*

Chenyu Zhang and Zhaslan Baraissov

*School of Applied and Engineering Physics,*

*Cornell University, Ithaca, NY 14853, USA*

Yu-Tsun Shao

*Mork Family Department of Chemical Engineering and Materials Science,*

*University of Southern California, Los Angeles, CA 90089, USA*

Cameron Duncan and Jared Maxson

*Department of Physics, Cornell University, Ithaca, NY 14853, USA*

Auralee Edelen

*SLAC National Accelerator Laboratory, Menlo Park, CA 94025, USA*

David A. Muller†

*School of Applied and Engineering Physics,*

*Cornell University, Ithaca, NY 14853, USA and*

*Kavli Institute at Cornell for Nanoscale Science, Ithaca, NY 14853, USA*

* Funded by the Center for Bright Beams, an NSF STC (NSF PHY-1549132).

† dm852@cornell.edu

‡ dm24@cornell.edu

## Abstract

Aberration-corrected Scanning Transmission Electron Microscopy (STEM) has become an essential tool in understanding materials at the atomic scale. However, tuning the aberration corrector to produce a sub-Ångström probe is a complex and time-costly procedure, largely due to the difficulty of precisely measuring the optical state of the system. When measurements are both costly and noisy, Bayesian methods provide rapid and efficient optimization. To this end, we develop a Bayesian approach to fully automate the process by minimizing a new quality metric, beam emittance, which is shown to be equivalent to performing aberration correction. In part I, we derived several important properties of the beam emittance metric and trained a deep neural network to predict beam emittance growth from a single Ronchigram. Here we use this as the black box function for Bayesian Optimization and demonstrate automated tuning of simulated and real electron microscopes. We explore different surrogate functions for the Bayesian optimizer and implement a deep neural network kernel to effectively learn the interactions between different control channels without the need to explicitly measure a full set of aberration coefficients. Both simulation and experimental results show the proposed method outperforms conventional approaches by achieving a better optical state with a higher convergence rate.

## I. INTRODUCTION

Aberration-corrected scanning transmission electron microscopes (STEM) use a series of multipole magnets to correct the intrinsic aberrations of the main round lenses in order to generate a sub-Ångström sized electron beam for atomic resolution imaging. Tuning the optics to achieve high resolution requires an accurate method of diagnosing the aberrations of the system and a way of adjusting each element to minimize these aberrations. The tuning procedure used on some modern corrected microscopes can take an experienced STEM user up to 2 to 3 hours to complete this preparatory process before high resolution can be achieved. Moreover, both experiment and theory have shown that the observed lifetime of a well corrected state is very short and suffers from intrinsic instability as a fundamental limit, necessitating constant tuning [1,2]. This further reduces the effective time available for targeted research objectives within a single microscope session.

The aberration function has been the standard language for characterizing the beam quality in an electron microscope. While some aberration coefficients (particularly those of low order) are straightforwardly related to particular misalignments of the beam, the coupling between different aberration orders and the indirect relation of high-order aberrations to microscope inputs makes this complex to use in practice. Existing methods for measuring the aberrations of the system are also time-consuming and inaccurate. In Part I of this paper, we showed that beam emittance growth, a widely used beam quality metric in accelerator physics, can be obtained in the context of electron microscopy by directly deriving it from the 2D electron wave function defined by the aperture function and the aberration function using the Wigner distribution [3]. The beam emittance is a single-valued metric, eliminating the necessity of dealing with individual aberration coefficients. Furthermore, we showed that a deep learning model can directly map the abundant phase space information embedded in electron Ronchigrams [4] to reliable estimates of beam emittance and that beam emittance is convex with respect to the aberration coefficients. Together, these properties indicate that beam emittance can serve as a black-box objective function to optimize as part of a fast and autonomous microscope tuning routine.

In Part II of this paper, we demonstrate the use of beam emittance minimization for electron microscope tuning. While we have replaced the collection of aberration coefficients with a single scalar objective, control of the microscope is still achieved through several (or, if full control of every optical element were used, dozens) of input channels. Optimization over such a high dimensional space with complicated couplings across dimensions is a difficult task. Bayesian optimization (BO) is a sequential search approach to global optimization of black-box functions that does not assume any functional forms, usually employed to optimize complex and expensive-to-evaluate functions. As a sample efficient and gradient free method, it preserves the greatest level of statistical rigor by incorporating prior knowledge from observed data to quantify uncertainty in the parameter space, which enables the balance between exploration and exploitation [5]. This capability is ideal for optimizing complicated scientific experiments where data collection is costly in time and resources and the outputs can be highly noisy and uncertain. Recent literature has shown success in using Bayesian optimization for experiment design and parameter tuning in microscopy [6], including automated aberration correction using a heuristic metric such as image variance [7]. In accelerator physics, BO has also been applied to tune free electron lasers (FEL) [8], laser wakefield accelerators [9] and laser-plasma accelerators [10], etc. Building upon generic Bayesian optimization, we further propose deep kernel Bayesian optimization (DKBO) for beam emittance minimization, which can converge faster with final states closer to the true optimum by learning the correlations between different input dimensions. In the setting of optimizing complex scientific instruments, this is particularly helpful as we still have full knowledge of the posterior to make sure the coupling between parameters is consistent with the physical laws. We have reported early results on this work in conference proceedings [4,11,12].

Here, we propose a fully automated scheme for autonomous online optimization of an electron microscope including the following steps: i) acquire an electron Ronchigram; ii) predict beam emittance growth using the deep learning model of Part I; iii) explore and update the microscope inputs using Bayesian optimization. A full diagram of this workflow is shown in Figure 1. We validate this unified machine learning approach to solving the optimization problem of aberration correction in electron microscopy with both simulation studies and real experiments and demonstrate that BO minimization of emittance growth is a rapid and effective method of aberration corrector tuning.

## II. METHODOLOGY

### A. From aberration function to beam emittance

We briefly review the derivations from part I. The standard expression to describe beam quality due to deviation from a perfect spherical focusing lens within the field of electron microscopy is the aberration function, where phase shift caused by imperfect lenses can be expanded as a polynomial in terms of the radial ($\alpha$) and azimuthal ($\phi$) angles using the Krivanek notation [13],

$$\begin{aligned} \chi(\alpha, \phi) &= \frac{2\pi}{\lambda} \sum_{n.m} \frac{C_{n,m} \alpha^{n+1} \cos\left(m\left(\phi - \phi_{n,m}\right)\right)}{n+1} \\ &= \frac{2\pi}{\lambda} \sum_{n,m} \frac{\alpha^{n+1}}{n+1} \Re\left[\tilde{C}^{*}_{n,m} e^{im\phi}\right] \end{aligned} \tag{1}$$

where $n$ is the order of aberration and $m$ the order of rotational symmetry.

At the same time within the field of particle accelerators, significant work goes into the preservation of beam brightness by minimizing the beam emittance growth within an accelerator. However there is a fundamental connection between the two metrics than can be made by describing the lens-induced aberrations on the electron beam in phase space using the Wigner-Weyl transform [3]. The statistical (root mean square) definition of emittance is adopted and shown to only include the gradient of the aberration function $\chi(\vec{\alpha})$.

$$
\begin{aligned}
\varepsilon_\chi^2 = &\quad \langle \nabla\chi^2 \rangle \langle \vec{\alpha}^2 \rangle - \langle \nabla\chi \cdot \vec{\alpha} \rangle^2 \\
= &\quad \left( \int d\alpha^2 A^2(\vec{\alpha}) |\vec{\nabla}\chi(\vec{\alpha})|^2 \right) \left( \int d\alpha^2 A^2(\vec{\alpha}) |\vec{\alpha}|^2 \right) - \int d\alpha^2 A^2(\vec{\alpha}) \vec{\alpha} \cdot \nabla\chi(\vec{\alpha})
\end{aligned} \tag{2}
$$

where $A(\vec{\alpha})$ is the aperture function defined in the angular basis with $\vec{\alpha}$. In part I of the paper, we derived properties of $\varepsilon_\chi^2$ that make it ideal for fast beam quality assessment including (i) independence of defocus and (ii) convexity in aberration coefficients.

Ideally, we can calculate the second moments in the first equality of Equation 2 by performing the integrals, However, it is usually more efficient to calculate the numerical gradient of the aberration function $\chi$ over a Cartesian grid, then obtain emittance from the determinant of the covariance matrix as discussed in Equation 1 of part I.

We emphasize that for the purpose of aberration correction we need not know the full emittance of the beam including finite source brightness, but only the extra emittance growth introduced by aberrations. In the rest of the paper, the two terms *emittance* and *emittance growth* are used interchangeably.

### B. General Particle Tracer (GPT) simulation

To test the application of Bayesian optimization for tuning of a corrected electron optical system, we perform ray tracing simulations of a simplified STEM with a hexapole aberration corrector. For simplicity while still capturing various orders of aberrations, we simulate a microscope consisting of 6 tunable elements, including 2 hexapole magnets (HP1, HP2) and 4 round transfer lenses (TL1, TL2, TL3, TL4). Details of the placement of optical components are listed in Table I. We use the General Particle Tracer (GPT) program for computing the electron trajectories through the simulated microscope. GPT is widely used in the design of charged-particle accelerators and beam lines [14] and simulates all 3D charged-particle dynamics in the system, including the higher order path deviations that give rise to aberrations.

For each simulation run, we compute the trajectories of 10,000 particles emitted from a point source located at the origin with a uniform angular distribution. The electrons travel along the optic axis through 2 condenser lenses and 1 adaptor lens with fixed excitation, followed by the remaining 6 tunable elements until the objective lens. A ray diagram is shown in Figure 3. We place a screen at the center of the objective lens to record the position and slope of each electron reaching the sample, from which the probe profile is generated based on the discrete simulated electrons via interpolation. We use a random phase plate to compute Ronchigrams from the simulated probe profile. The random phase plate was tuned with the optimized size and Gaussian blurring to mimic the scattering by a real amorphous phase object, and the same phase plate is used between subsequent runs. We normalize the emittance growth and defocus by their maximum values as labels for the CNN.

TABLE I: Optical components simulated in GPT (in order)

| Optical component | Location ($z$ (m)) |
|---|---|
| Condenser 1 | -0.36 |
| Condenser 2 | -0.32 |
| Adaptive lens | -0.24 |
| Hexapole element 1* | 0.0 |
| Transfer lens 1* | 0.06 |
| Transfer lens 2* | 0.18 |
| Hexapole element 2* | 0.24 |
| Transfer lens 3* | 0.5 |
| Transfer lens 4* | 0.72 |
| Objective lens | 0.99 |

* Tunable elements being optimized.

### C. Bayesian optimization

Bayesian optimization is a sampling efficient and gradient free method suitable for cases where the objective is a black-box function and evaluations are expensive. It makes an assumption on the prior distribution of the observed inputs and update toward the next point by optimizing the model with respect to a certain acquisition function, usually derived from the posterior. This feature enables balance between exploration and exploitation and much faster convergence than traditional gradient-based methods.

The root of Bayesian optimization stems from Bayes rule. We want to optimize a function $f: \mathcal{X} \rightarrow \mathbb{R}$ where $\Omega$ is the set of input parameters we want to search over. While $f$ is expensive to evaluate, we can model $f$ as a probability distribution. Given observed data $D = \{\mathbf{x_1}, \mathbf{x_2}, \dots, \mathbf{x_D}\}$, the posterior predictive distribution of observing $f(\mathbf{x})$ for new variable $\mathbf{x}$ is given by the conditional probability,

$$\mathbf{P}\big(f(\mathbf{x}) \mid f(\mathbf{x_1}), f(\mathbf{x_2}), \ldots, f(\mathbf{x_D})\big) \tag{4}$$

With this conditional distribution, we can (i) estimate $f(x)$ for values of $x$ we have not observed yet, and (ii) choose the next value of $x$ we want to compute as the optimization proceeds.

The special case of assuming a Gaussian process (GP) prior over $D$ makes the above conditional distribution tractable in closed form. A GP model assumes an infinite collection of random variables such that each finite set has a joint Gaussian distribution specified by its mean and covariance, or kernel function.

$$f(\mathbf{x}) \sim \mathrm{GP}[\mu(\mathbf{x}), k(\mathbf{x}, \mathbf{x})] \tag{5}$$

where we get to choose the hyperparameter for the kernel to capture the covariance between different inputs.

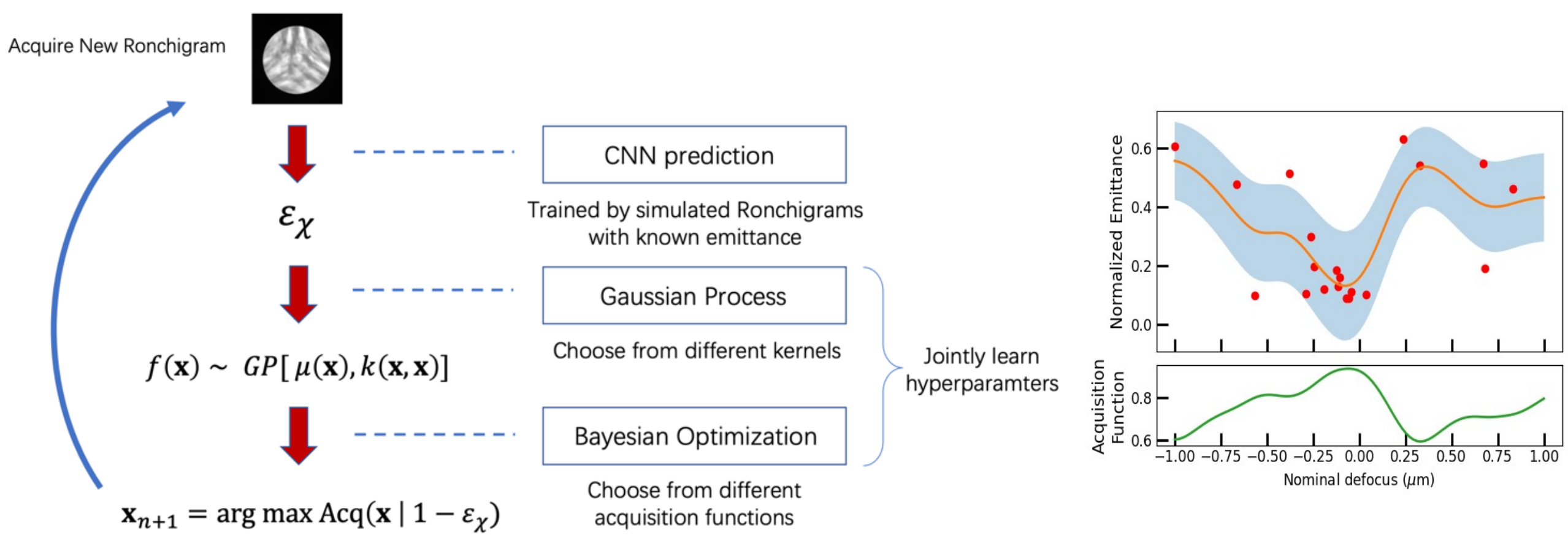

FIG. 1: Workflow of online optimization of an electron microscope.

## C. Physics-informed kernel

In our case, the objective function is beam emittance growth, and the input parameters are ideally values of lens channels, but more straightforwardly the aberration buttons provided by the corrector manufacturers, which are calibrated to a stationary mapping of lens currents[1]. However, since the experimental aberration function is unknown, we cannot obtain emittance directly in experiments using Equation 2. In part I of the paper, we showed that a deep learning model can build a mapping from experimentally accessible electron Ronchigrams, which becomes the objective function we optimize here $\{\boldsymbol{x} \in \mathbb{R}^d, \varepsilon_\chi\}$.

The choice of kernel is key to capturing correlations between inputs in the GP. The mostly popular one is the radial basis function (RBF) kernel, which maps the inputs to an infinite dimensional space. Other popular kernels are discussed in Appendix B.

$$k_{\mathrm{RBF}}(\mathbf{x_1}, \mathbf{x_2}) = \exp\left(-\frac{1}{2}(\mathbf{x_1} - \mathbf{x_2})^{\top}\Theta^{-2}(\mathbf{x_1} - \mathbf{x_2})\right) \quad (5)$$

However, most generic kernels like this are isotropic, where the kernel function is only dependent on the distance between each pair of inputs. In other words, Θ is a diagonal matrix. We can further improve the model by leveraging our knowledge of the physical system, i.e., we clearly know the inputs of the microscope are coupled by magnetic fields in some way, which would constitute the off-diagonal elements. Now the question to answer is how to extract the correlations.

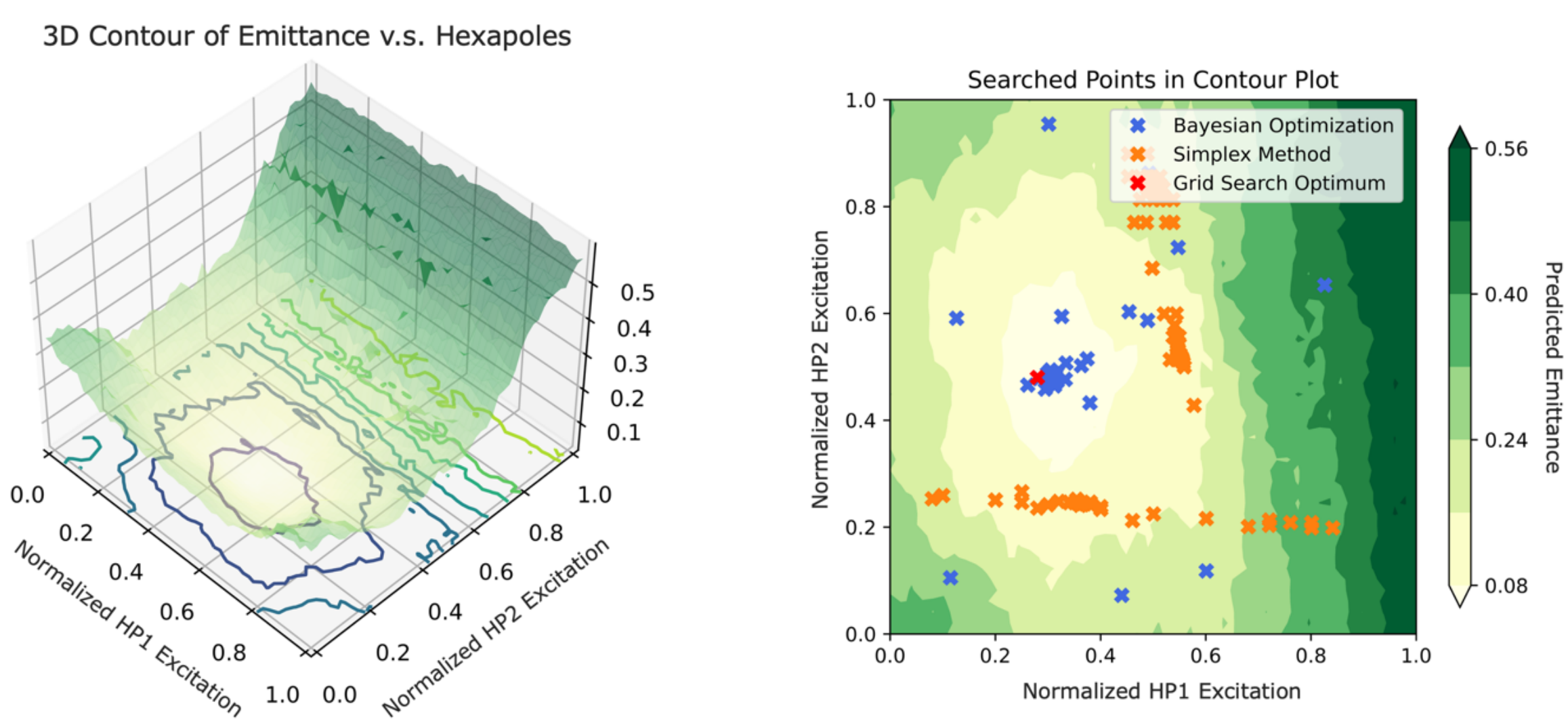

FIG. 2: Left: 3D contour of predicted emittance growth v.s. hexapole inputs. Right: Searched

[1] Most manufacturers offer a certain level of access to the components in the microscope. However, direct control over the magnetic coils is usually not straightforward. Sometimes it is possible to control the estimated aberration coefficients instead.

points by Bayesian optimization v.s. Nelder Mead Simplex method.

In some cases, we precisely know the behavior of the objective function with certain inputs. For example, in part I of the paper we have proved that beam emittance growth of an electron microscope is a convex function of aberration coefficients and the Hessian of $\varepsilon_\chi^2$ with respect to the aberration coefficients is a positive semidefinite matrix.

$$\begin{aligned} H_{n,n',m} &= \frac{d^2\varepsilon_\chi^2}{d\tilde{C}_{nm}d\tilde{C}_{n'm}} \\ &= \alpha_0^{n+n'+2}\left(\frac{1}{n+n'+2}-\frac{4\delta_{m0}}{(n+3)(n'+3)}+\frac{m^2}{(n+1)(n'+1)(n+n'+2)}\right) \end{aligned} \quad (6)$$

In this case, we can easily extract the correlation matrix between input dimensions from the Hessian by a transformation $h = \text{diag}\,(H)^{-\frac{1}{2}}H\,\text{diag}\,(H)^{-\frac{1}{2}}$ [8], and integrate the length-scale factors by replacing $\Theta$ with $\Theta^{\frac{1}{2}}h\Theta^{\frac{1}{2}}$.

Nevertheless, the above nice property of the objective function with respect to input parameters is a rare case. In practice, the ultimate control of the microscope is determined by actual electric currents running in the aberration corrector coils. Those inputs are usually coupled, nonlinear and build hysteresis in complex ways [1,2]. Sometimes it is possible to gain insight by simulations. For example, in Figure 2 we plot the 3D contour of predicted beam emittance growth with respect to the excitation of each of the two hexapole elements in the simulated microscope. One can roughly visualize the convexity in local areas. However, we note that this approach is not necessary when lower dimensional tasks are simple enough to be handled by regular Bayesian optimization, and quickly becomes prone to bias and infeasibility as the number of dimensions involved increases. Therefore, we design and provide a simpler but more general solution in the next section.

### D. Deep kernel Bayesian optimization

Deep kernel learning [15] is a concept that combines a deep neural network (DNN) with a standard GP and base kernel to outperform standard GPs by utilizing the expressive power of DNNs to learn finer representations in the high dimensional space. Specifically, the model aims to transform the inputs (predictors) $\boldsymbol{x}$ in a base kernel $k\,(\boldsymbol{x}_i, \boldsymbol{x}_j \mid \boldsymbol{\Theta})$ with hyperparameters $\boldsymbol{\Theta}$ to an embedding space by

$$k(\mathbf{x}_i, \mathbf{x}_j \mid \mathbf{\Theta}) \rightarrow k(g(\mathbf{x}_i, \mathbf{w}), g(\mathbf{x}_j, \mathbf{w}) \mid \mathbf{\Theta}, \mathbf{w}) \tag{7}$$

where $g\ (\boldsymbol{x}, \boldsymbol{w})$ is a non-linear mapping given by a deep architecture, such as a DNN, parametrized by weights $\boldsymbol{w}$. It is shown one can jointly update the DNN weights $\omega$ and the GP hyperparameters $\Theta$ by optimizing the log likelihood with gradient descent [15]. A natural thought would be: can we directly replace GP in the Bayesian optimization with the more powerful DKL? Despite its success, it is only until recently that an increase interest in combining DKL and BO starts to appear. For example, deep kernel learning is shown to do effective transfer learning for few-shot BO [16]. A more recent paper aims to build a more general deep kernel Bayesian optimization framework and shows that together with Monte Carlo dropout layers it serves equivalently as an approximate posterior sampling method [17].

In our specific use case, as we proceed and acquire data pairs of input lens parameters and corresponding emittance growth, the deep kernel weights are concurrently optimized. We expect deep kernel learning to extract the complex coupling between different lenses and/or different orders of aberrations in order to achieve faster convergence.

## III. RESULTS

### A. Optimization of a simulated microscope

We start the validation of our method with simulations studies using the ray tracing simulations introduced in section II and conduct the Baysian optimization following the framework outlined in Figure 1.

Figure 3 (a) shows the ray diagram of the simulated microscope before optimization (a random initialization) and after optimization of HP1 and HP2. At first glance we can tell the optimized ray diagram preserves more axial symmetry. Further inspecting the corresponding Ronchigrams before and after the optimization in Figure 3 (b), we find that after optimization the Ronchigram has shown a much larger flat area at the center, indicative of reduced phase distortions. This is verified by visualizing the actual spatial distribution of electrons along the longitudinal direction at the screen in Figure 4. In the uncorrected case (left panel), the distribution is of rotational symmetry of order of 3 both at the core and the periphery, which is typically from 3-fold astigmatism. In the corrected case (right panel), the center is almost perfectly round while only at the periphery does the distribution show rotational symmetry of order of 6. This indicates the two hexapoles have been optimized to counteract each other to correct the aberrations of the objective lens. The sixfold astigmatism of the simple design of hexapole corrector simulated here cannot be nulled without the addition of further multipole elements [18]. Since the particles acquire angular momentum inside the solenoids, x and y coordinates are coupled and cannot be analyzed separately. Instead, we plot the phase space distribution in polar (angular) coordinates in Figure 5. We can obtain an estimate of the beam emittance using the root-mean-square definition. The optimization has indeed reduced emittance growth. The phase space area occupied by the beam has shruk after optimization, particularly for electrons with angles below ~30 mrad which correspond to the enlarged flat area of the Ronchigram. Electrons arriving at greater angles still posess substantial position deviations owing to higher order aberrations which remain uncorrected. Optimization of this simulated microscope with a Bayesian optimizer converges in 40 iterations on average which translates to less than 5 minutes on a commercial desktop with a 24-core CPU and 124 GB RAM. This time is dominated by the GPT simulation, which relies on the CPU only, rather than the evaluation of the Bayesian optimization .

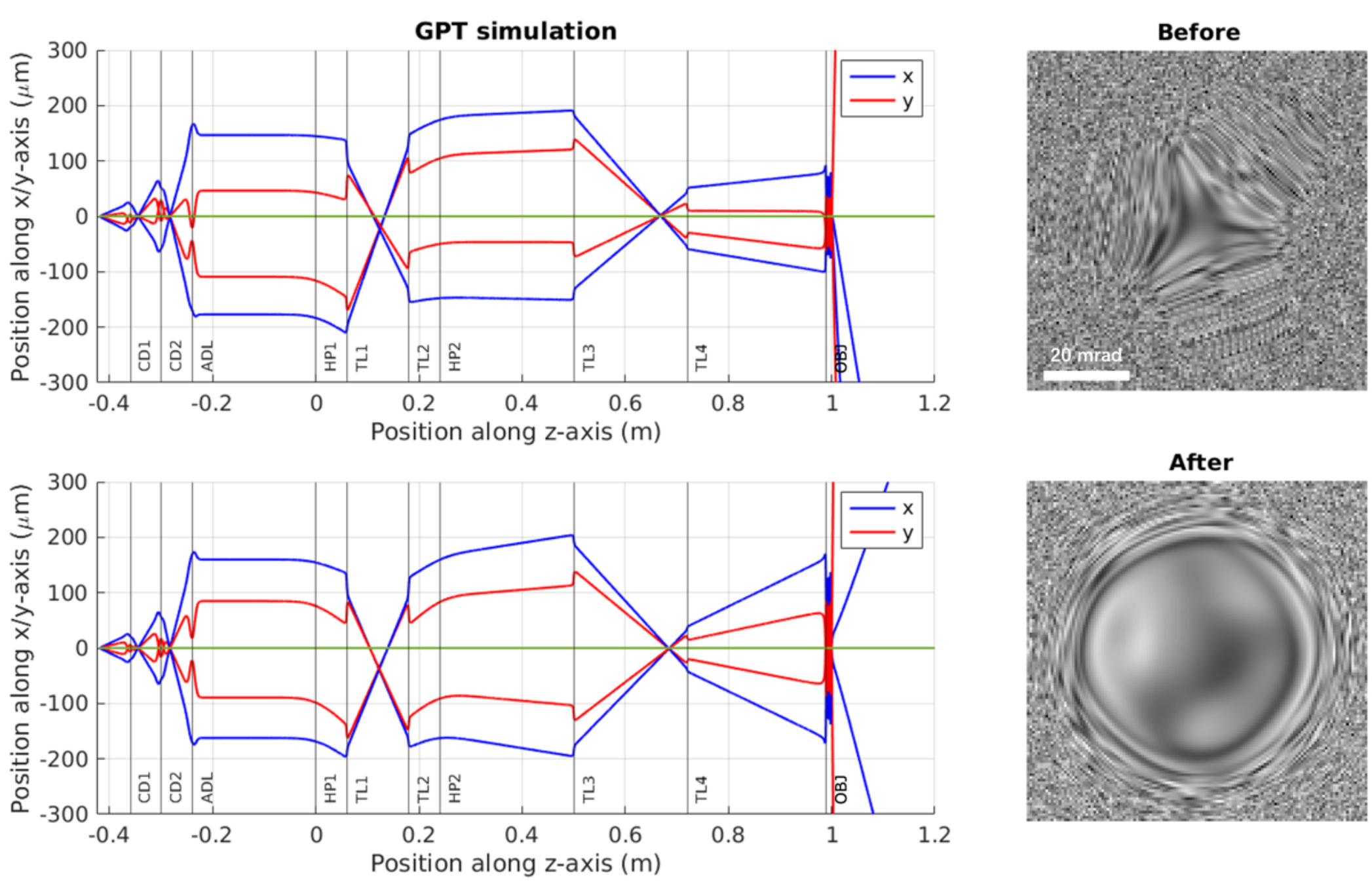

(a) (b)

FIG. 3: (a). Ray diagram of the simulated microscope after and after optimization, with the locations of the condenser lenses (CD), hexapole elements (HP), transfer lenses (TL), adapter lens (ADL), and objective lens (OBJ). Pairs of axial rays, corresponding to particles initially emitted from the origin with slope along the $x$ and $y$ global coordinate axes, are shown. Top: before optimization; bottom: after optimization. (b). Simulated electron Ronchigrams. Top: before

optimization of the hexapole elements; bottom: after optimization of the hexapole elements. The Ronchigram after optimization shows a much larger flat area in the center, corresponding to a larger usable angular range with low optical distortions.

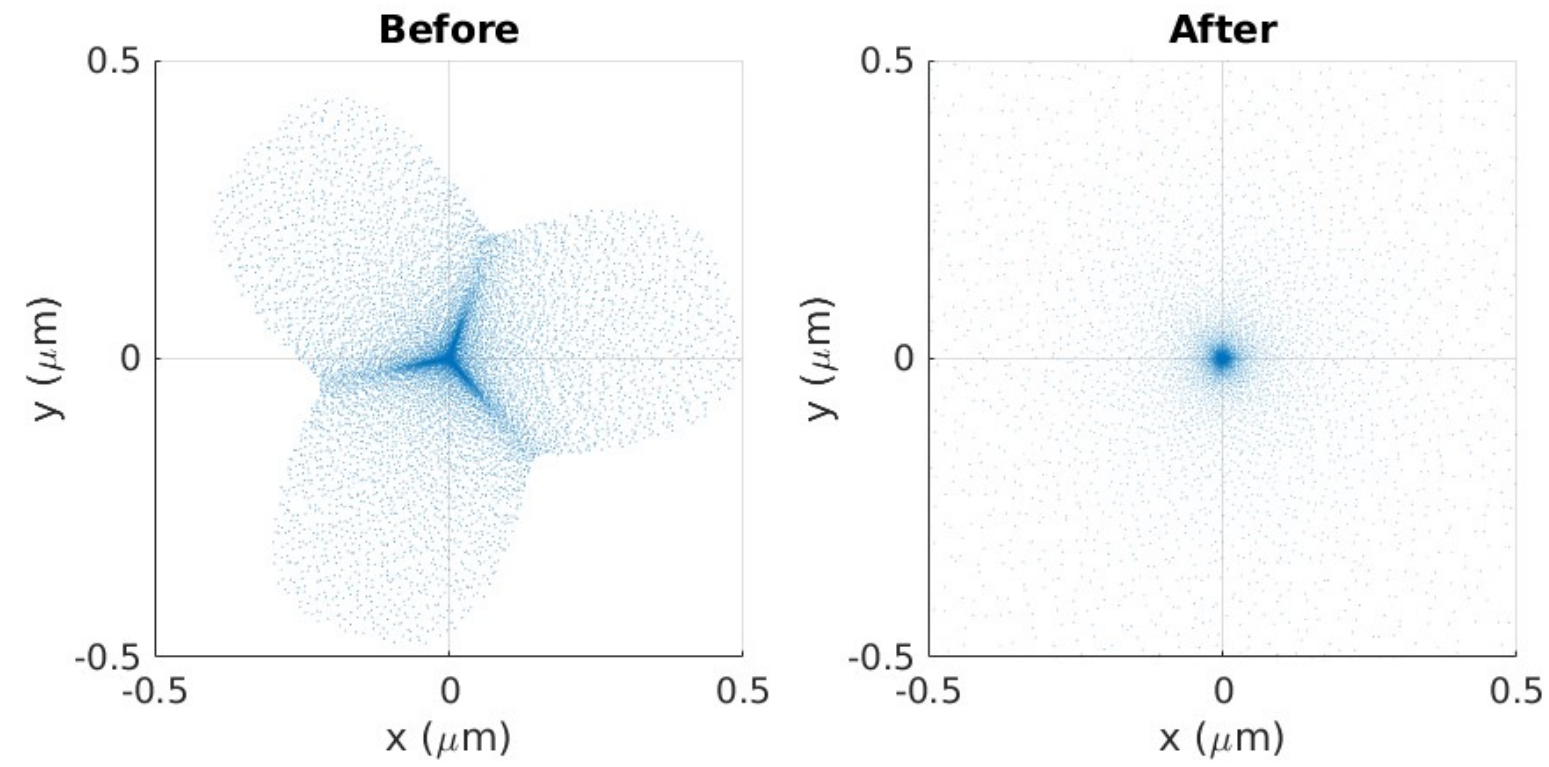

FIG. 4: Transverse distribution of electrons at the center of the objective lens. Left: before optimization of the hexapole elements, beam FWHM = 0.3850 $\mu$m; Right: after optimization of the hexapole elements, beam FWHM = 0.0064 $\mu$m. The comparison indicates that optimization of aberration correctors according to the minimization of beam emittance can effectively eliminate lower order aberrations.

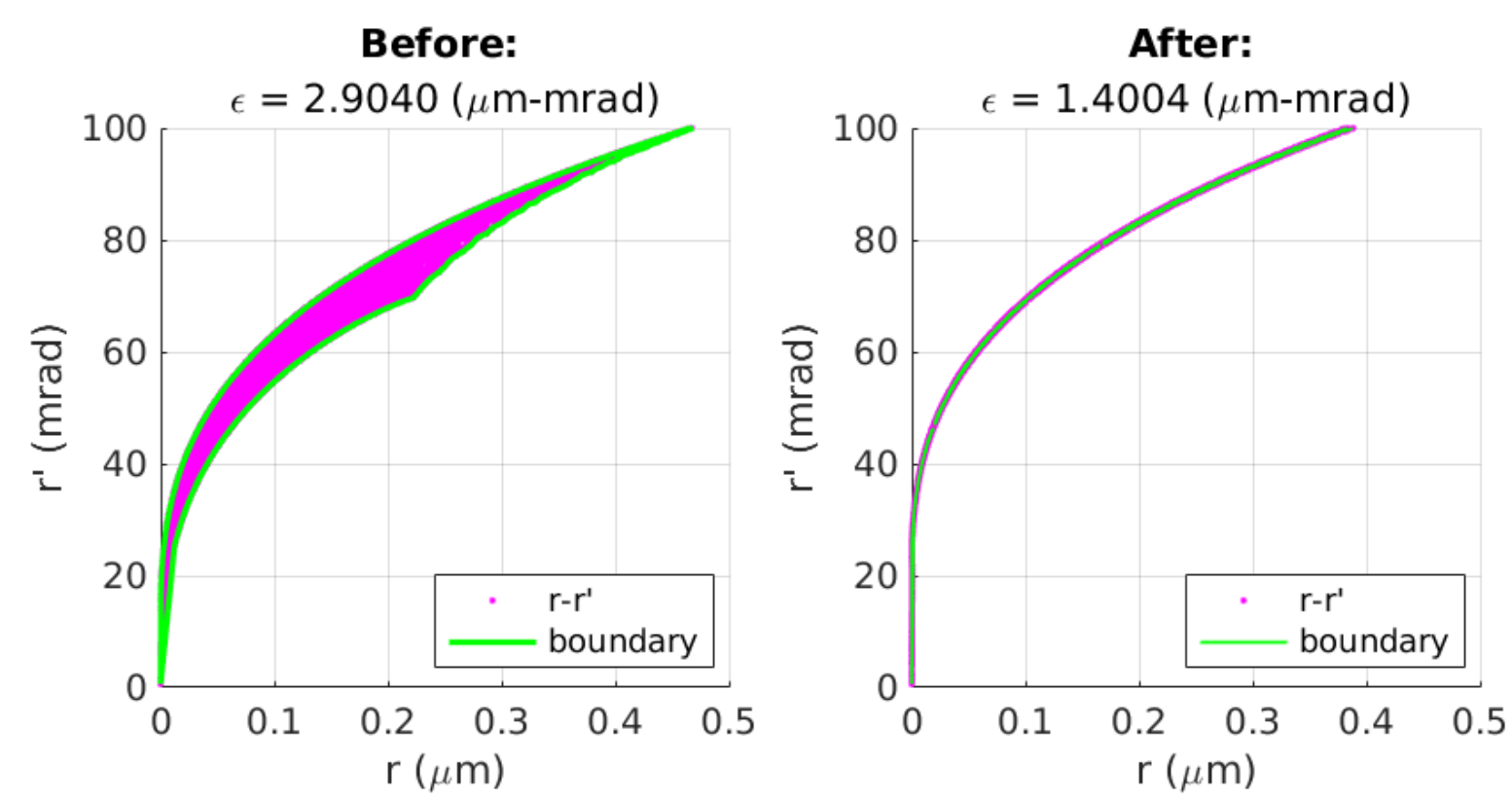

FIG. 5: Phase space distribution of the electrons at the center of the objective lens in polar coordinates. Left: before optimization of the hexapole elements; Right: after optimization of the hexapole elements.

We next benchmark the convergence behavior of BO. We initialize with 10 randomly generated Ronchigrams by GPT to construct the first GP. The parameter space is explored and exploited according the choice of kernel and acquisition function. In order to test the ability of the different kernels to learn and exploit the coupling between input parameters, we now optimize both the four transfer lenses as well as the hexapole strengths. We test the generic RBF kernel, the Matern kernel, and the deep kernel with an upper confidence bound (UCB) acquisition function with hyperparameter setting $\beta = 0.2$. For comparison, we also perform the optimization using the Nelder

Mead Simplex method as a baseline, as it is still the major method widely used in electron optical design. We run 20 independent repetitions to benchmark the performance. The budget of iterations for each run is set to be 300 for the GPT-6D case (optimization of all six tunable elements: HP1, HP2, TL1, TL2, TL3, TL4). More benchmark results with other choices of kernels and hyperparameters are discussed in Appendix C.

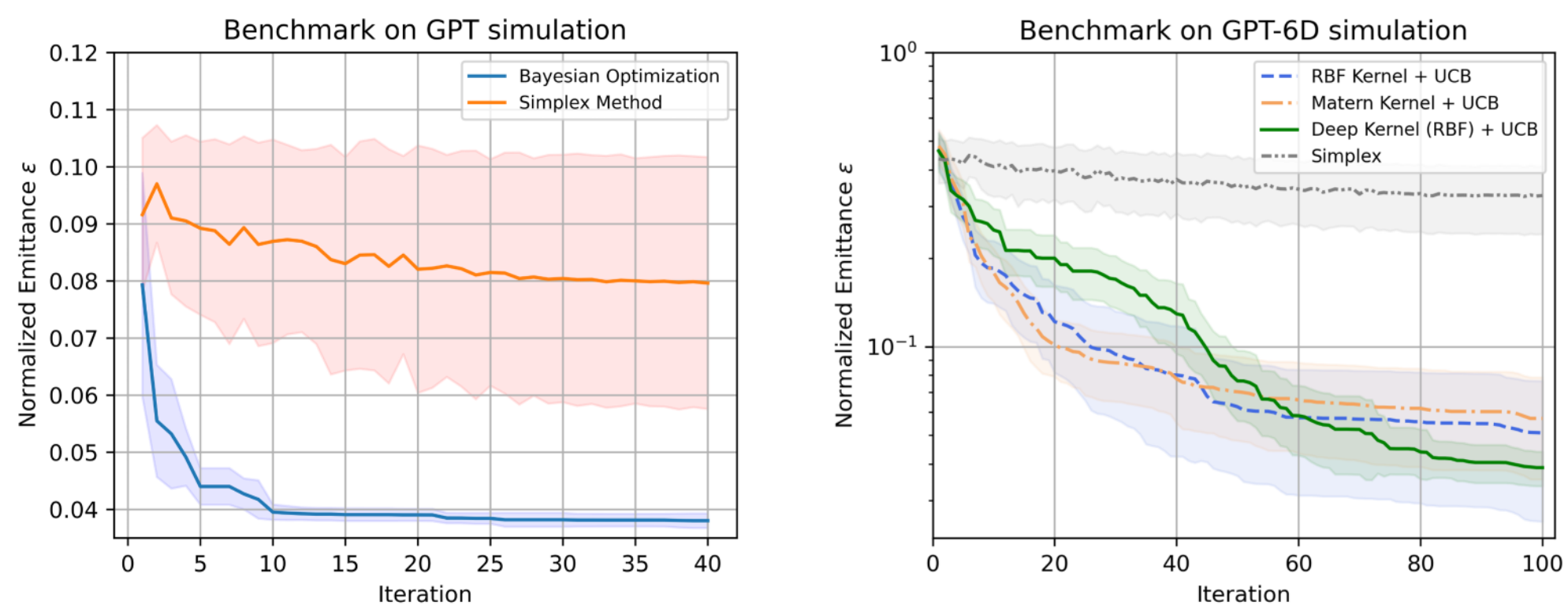

FIG. 6: Left: The simple case of 2D (hexapoles-only) optimization. Bayesian optimization significantly outperforms the Simplex method, converging within 10 iterations. Right: The more complex case of 6D optimization. Benchmark results of DKBO v.s. generic BO v.s. Simplex method on the GPT-6D simulation within 100 iterations.

Figure 6 shows that Bayesian optimization always outperforms the Simplex method regardless of the choice of kernel. Interesting results arise between the RBF/Matern kernel and the deep kernel (green solid line). We notice that DKBO outperforms in the long term, though underperforms in the beginning. This can be explained by the overfitting of a DNN. Essentially deep kernel trains a DNN on observed data points to extract the underlying lower-dimensional representation. When the number of observations is small, the DNN overfits. However, as observations accumulate, we start to benefit from the expressive power of the DNN, which help us reach a lower emittance at higher convergence rate. This is further justified by the narrower standard error of the DKBO curve, which indicates the changeover from exploration to exploitation.

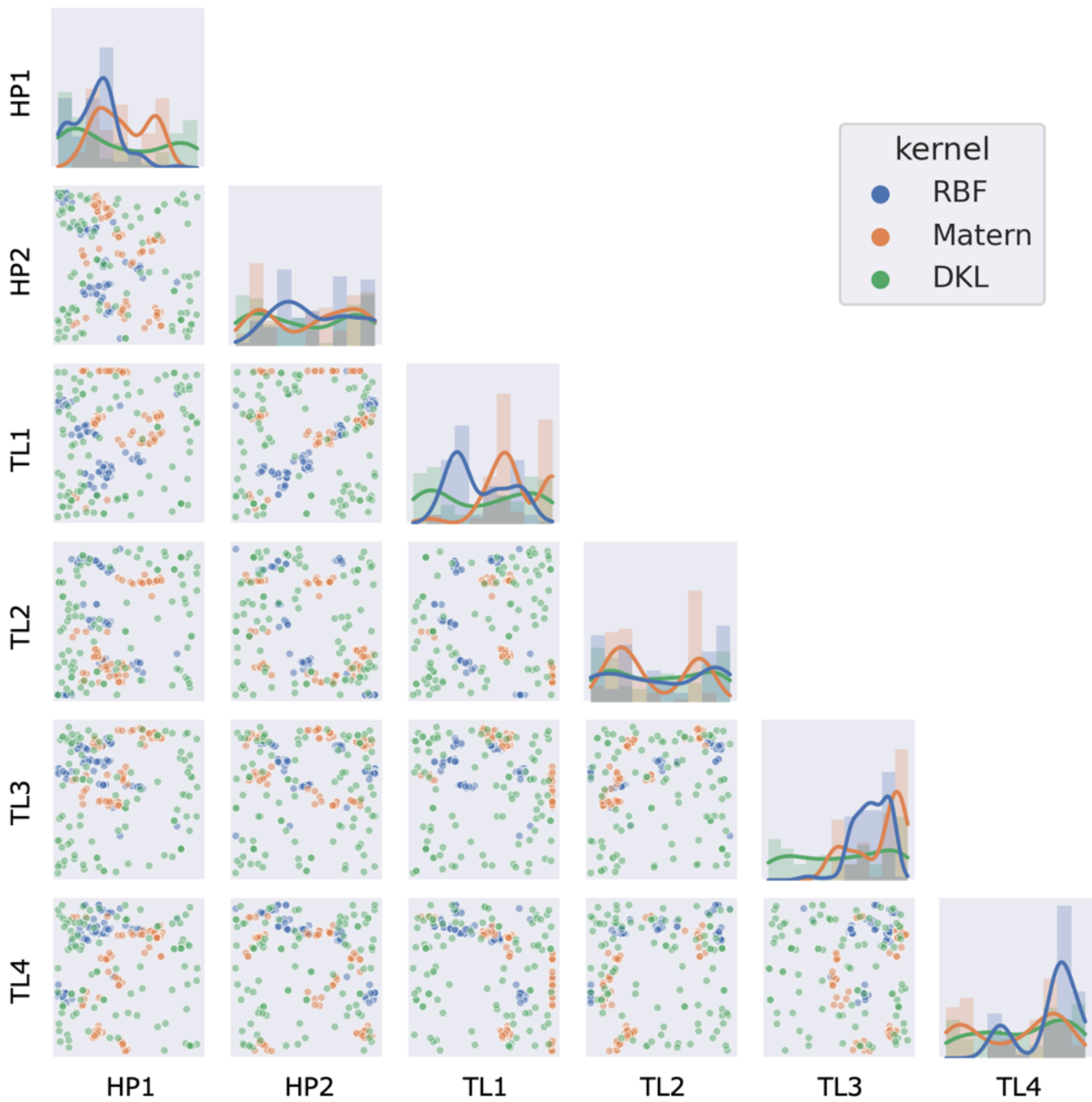

FIG. 7: Distribution of explored points in the parameter space within 100 iterations by different BO kernels: RBF, Matern and DKL. Each grid represents a 2D projection of the parameter space onto a pair of parameters being optimized, with the colored scatter plots illustrating the points queried by BO. The goal is to explore all of the parameter space efficiently, which DKL does by extracting correlations between parameters even when far apart. The isotropic RBF and Matern kernels, on the contrary, are trapped more in localized states.

In Figure 7 we further show how deep kernel learning is capable of extracting the underlying correlations between input dimensions even at far distances in the parameter space. The deep kernel explores the hyperparameter space more aggressively and eliminates the chance of being trapped at local optima while the isotropic RBF and Matern kernels can easily localize.

### B. Online optimization of real microscopes

We next test our Bayesian optimization approach on real electron microscopes, including the ThermoFisher Titan Cryo-S/TEM and the ThermoFisher Spectra 300, equipped with the CEOS D-CORR aberration corrector and the CEOS S-CORR aberration corrector respectively.

Control of the ThermoFisher microscope is made possible by the AutoScript TEM Software[2] built in Python 3. The aberration correctors are controlled via an RPC protocol provided by CEOS. The CEOS RPC protocol only provides access to aberration correction commands instead of direct lens channels, e.g. A1, A2, B2, C3, with both x and y axis and either "coarse" or "fine" option. These aberration correction commands are the same as user commands in the corrector graphical interface. We emphasize that ultimately the only inputs needed to align the corrector are the input current values of lens channels. However, due to the lack of access, we demonstrate optimizing using these aberration buttons, which would yield equivalent results given that the factory calibrated mapping between the two is stationary. The "coarse" and "fine" versions of the controls in the CEOS software do not refer to different step sizes of the correction, and in fact generally use completely different coils and so have different higher-order side effects. In addition, we implement a "deGauss" command to account for hysteresis effects, which sets all lens channel values to zero for two seconds after every 5 iterations in the BO.

Figure 8 presents the results of online tuning of the ThermoFisher Titan Cryo-S/TEM and the ThermoFisher Spectra 300 microscopes with our proposed Bayesian optimization. Similar to the simulation, we first acquire 10 Ronchigrams with random inputs and predict their emittance growth values with the CNN, upon which we initialize the GP for Bayesian optimization. After 50 iterations all Ronchigrams show a much larger flat area at the center, which indicates greatly improved probe quality. The average time for 50 iterations is 4 minutes, mostly limited by the acquisition time of single Ronchigrams and the optional "deGauss" command. The final normalized emittance values measured by the CNN after Bayesian optimization were 0.0266 ± 0.0037 (Cryo S/TEM) and 0.0264 ± 0.0045 (Spectra 300) averaged across 10 runs.

For comparison, we also tuned the aberration corrector of the Spectra 300 several times using the standard Zemlin tableau method implemented in the corrector software. Each Zemlin tableau measurement takes roughly 3 minutes, and it usually requires several runs to achieve reasonably well corrected state. This approach takes the extra effort to measure each individual aberration coefficient of the Zernike polynomial aberration function using a tilt series of STEM images at the expense of much longer acquisition time. Figure 9 presents the aberration measurements across several iterations of tuning with the Zemlin tableau. While the aberration coefficients do largely come closer to the center (zero values), the precision is low. These uncertainties in the polynomial coefficients translate to phase space spread, reflected as a residual emittance growth. Finally, we examine the Ronchigrams acquired from the corrected states and predict final emittance with the same trained deep neural network 0.0977 ± 0.0041 (Spectra 300), averaged across 10 runs. Therefore, our Bayesian approach outperforms existing Zemlin-like methods both in terms of speed and final convergence. Further test results using a Nion UltraSTEM microscope are provided in Appendix A, but a full test was not possible as the instrument was decommissioned during the course of this project.

---

[2] AutoScript: https://www.thermofisher.com/us/en/home/electron-microscopy/products/software-em-3dvis/autoscript-tem-software.html

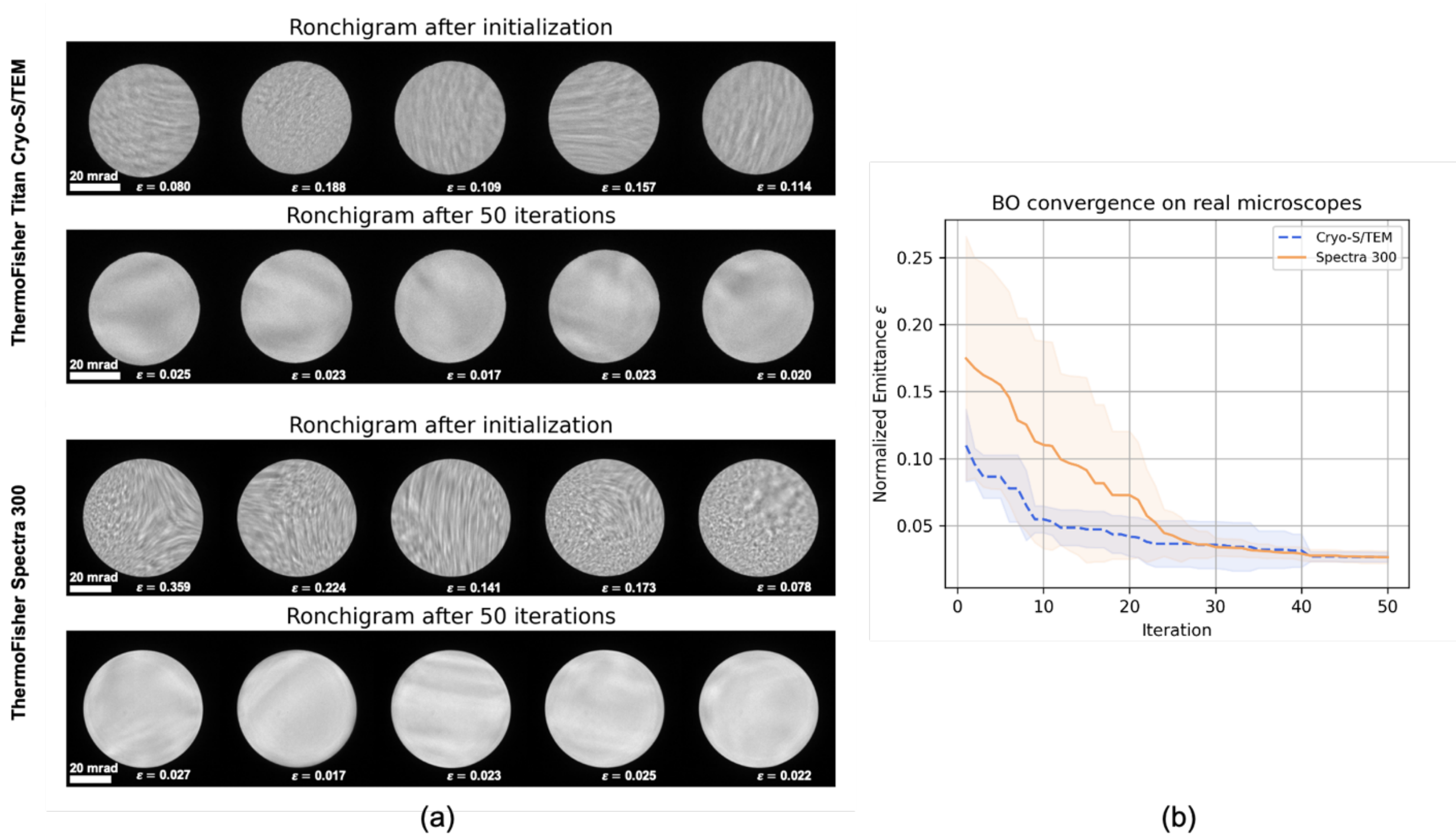

FIG. 8: (a) Ronchigram appearance before and after 50 iterations with online Bayesian optimization of the Titan Cryo S/TEM (top) and Spectra 300 (bottom). (b) Convergence behavior of BO on the two microscopes. Final Ronchigram emittance estimates for 10 runs were 0.0266 ± 0.0037 (Cryo S/TEM) and 0.0264 ± 0.0045 (Spectra 300). The average time for 50 iterations is 4 minutes, mostly limited by the acquisition time of a single Ronchigram and the deGaussing of the lenses.

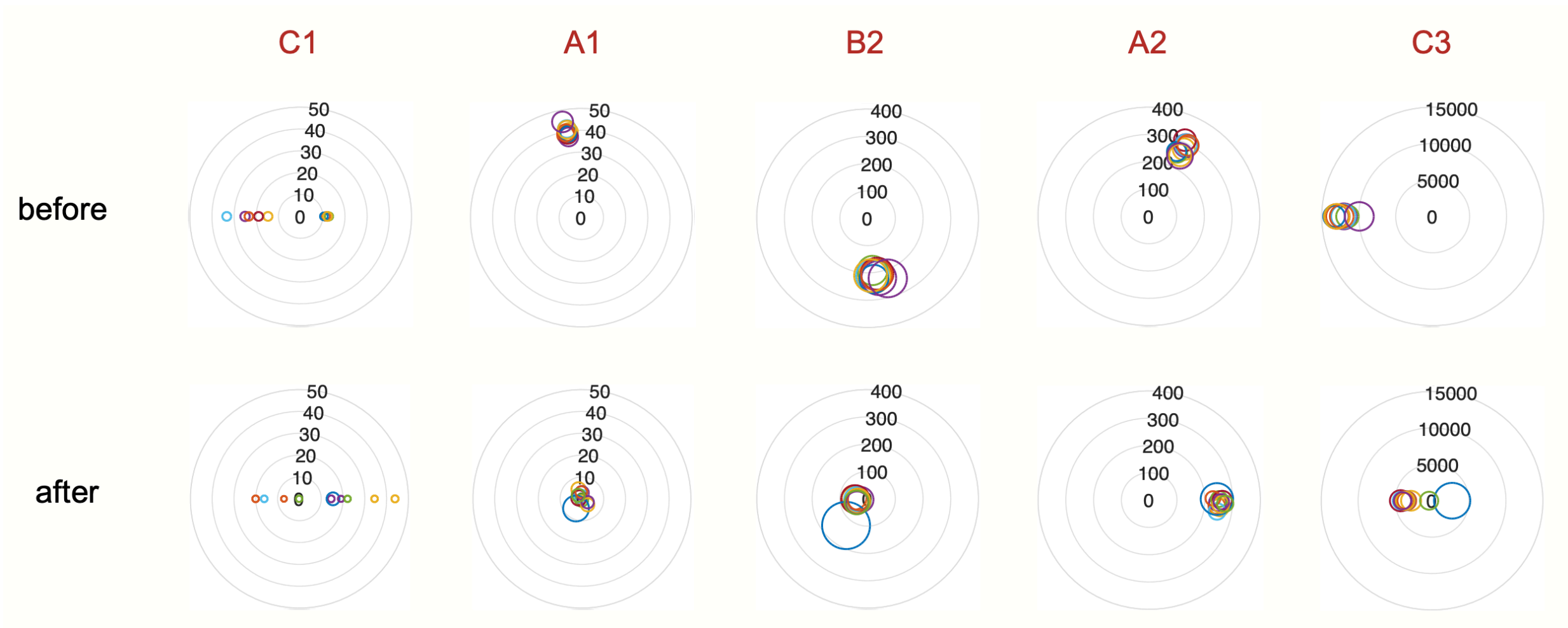

FIG. 9: In comparison, we show the aberration measurements from the Zemlin Tableau method before and after aberration correction with 95% confidence intervals (unit: nm). While the measured values become closer to zero after correction, the scatter in the measurements is still quite large. Data is plotted in polar coordinates to reflect angles and is collected from the CEOS S-Corr corrector software on a Spectra 300. Final Ronchigram emittance estimates for 10 runs: $0.0977 \pm 0.0041$.

## IV. DISCUSSION

Conventional aberration correction software typically focuses on estimating and minimizing individual aberration coefficients. This process involves correlating segmented images either in real space or reciprocal space, which is time consuming due to the large number of images that must be acquired for each measurement. In part I of this paper, we have demonstrated that this traditional approach can be greatly simplified by leveraging the mathematical equivalence between aberration correction and the minimization of beam emittance. Furthermore, we show one can train a convolutional neural network (CNN) to extract the beam emittance growth through the column from the electron Ronchgrams directly. Consequently, the correction process does not need to involve the estimation of multiple aberration coefficients separately; rather, it can be reduced to optimizing a single-valued metric with the same set of parameters, streamlining the entire optimization process.

The conventional approach for aberration correction is the Zemlin tableau, where each individual aberration coefficient is estimated by regression on illuminations of the sample at different beam tilts. In this paper, we compare our approach against it via online optimization of real microscopes, ThermoFisher Titan Cryo-S/TEM and ThermoFisher Spectra 300. Our findings demonstrate that Bayesian optimization achieves rapid convergence within approximately 50 iterations, reducing the total tuning time to around 4 to 5 minutes. In contrast, the conventional

Zemlin Tableau method requires roughly 3 minutes to complete one round of aberration measurement, and multiple repeated measurements are required throughout the tuning process. This results in considerably more time expenses. As pointed out in the results section, Zemlin Tableau can also struggle to stabilize and achieve consistent measurements of individual aberration coefficients. Uncertainty in the aberration measurements leads to incorrect adjustments to the lens currents, and leading to the tendency of the correction to "overshoot," particularly for higher order aberrations. While the objective is to improve beam quality, the overall uncertainty in beam quality evaluation can add up proportionally with errors in each coefficient measurement, underscoring the advantage of our Bayesian optimization approach.

An alternative approach for bypassing the measurement of individual aberration coefficients is to use an image contrast based metric for optimization. While this is a heuristic metric not directly derivable from the probe function, the calculation is fast and does not require additional training for the model. Existing STEM auto-tuning programs (such as Sherpa on ThermoFisher microscopes) use a brute-force search over the values of the low-order corrections to optimize this metric. Recent work by Pattison et al. [7] improves upon this approach by using Bayesian optimization to accelerate the procedure, and demonstrated convergence for all lower order aberrations faster than by using the tableau method. Using image contrast as the optimization metric requires a stable, high-contrast and radiation-resistent sample during tuning, while our Ronchigram-based approach requires an amorphous sample area. This can become beneficial for aberration correction on the fly in the middle of an experiment.

## V. CONCLUSION

In this paper, we extend our work from part I and introduce an online Bayesian optimization technique tailored to the task of aberration correction in electron microscopy. This method effectively optimizes the electron probe by minimizing beam emittance growth—a single-valued metric—and simultaneously incorporates correlations among input parameters into the full posterior distribution of the objective function. This allows the algorithm to not only achieve rapid convergence but also maintain a physical understanding of the system dynamics.

The ultimate aim of our research is to develop a fully automated scheme capable of performing online tuning of an electron microscope, effectively taking over tasks traditionally managed by human operators. By significantly reducing the tuning time from hours to mere minutes, our method demonstrates a transformative improvement in efficiency and accuracy. We validate our approach by testing it on three state-of-the-art aberration-corrected electron microscopes: the ThermoFisher Cryo S/TEM, the ThermoFisher Spectra 300, and the Nion UltraSTEM. The results confirm that our approach is not only feasible but also highly efficient in practical applications.

Future work will focus on deeper integration between software and hardware components. This will involve direct communication with the optical components of electron microscopes, ideally through collaboration with manufacturers. This work may have applications beyond electron microscopy; the framework presented is adaptable to a variety of optimization tasks across different domains of science and engineering, offering a generalizable strategy for high-dimensional optimization problems.

**ACKNOWLEDGMENTS**

We thank John Grazul and Mariena Sylvestry Ramos for technical support and maintenance of the electron microscopes. We thank Heiko Müller and his colleagues at CEOS for helping us access their corrector optics. We thank Adi Hanuka for helpful discussions and guidance on the implementation of Bayesian optimization. We thank Alexander Pattison for with the CEOS RPC gateway wrapper. This work was funded by the Center for Bright Beams, an NSF STC (NSF PHY-1549132). Electron Microscopy facilities support from the Cornell Center for Materials Science, an NSF MRSEC supported by DMR-1719875, and the PARADIM Materials Innnovation Platform (NSF DMR-2039380)

**Appendix A: Online optimization of the Nion UltraSTEM**

With the Nion UltraSTEM, control of the microscope is made possible by Nion Swift [3]. The software enables access to the camera acquisition and aberration corrector controls up to the second order. Similarly to the simulation, we first acquire 10 Ronchigrams and predict their emittance growth values with the CNN, upon which we initialize the GP for Bayesian optimization. Details of the CNN training can be referred to part I of the paper and Appendix D. Figure A1 presents the results of online tuning the Nion UltraSTEM microscope with our proposed Bayesian optimization. The Nion microscope used a larger aperture size of 45 mrad which would include more higher order aberrations in the Ronchigram, but also had some intrinsic limitations in the lens design that inhibited perfect tuning even at smaller angles.

[3] Nion Swift: https://nionswift.readthedocs.io/en/stable

Since Nion Swift is implemented in a way to record absolute values of tuning, in Figure A2 we can plot the queries of individual aberration coefficients and the corresponding emittance growth at each iteration in a single run. The search quickly stabilizes after roughly 20 iterations and almost localizes after 40 iterations. This is consistent with the results on ThermoFisher microscopes in Figure 8.

**Appendix B: Other Kernels**

The kernel plays a significant role in the modeling of Gaussian processes, which encodes the covariance between each pair of the inputs. The go-to choice of kernel is the radial basis function (RBF) kernel. However, we further discuss how we can leverage our knowledge of the physics in the microscope system to choose more suitable kernels and improve existent kernels.

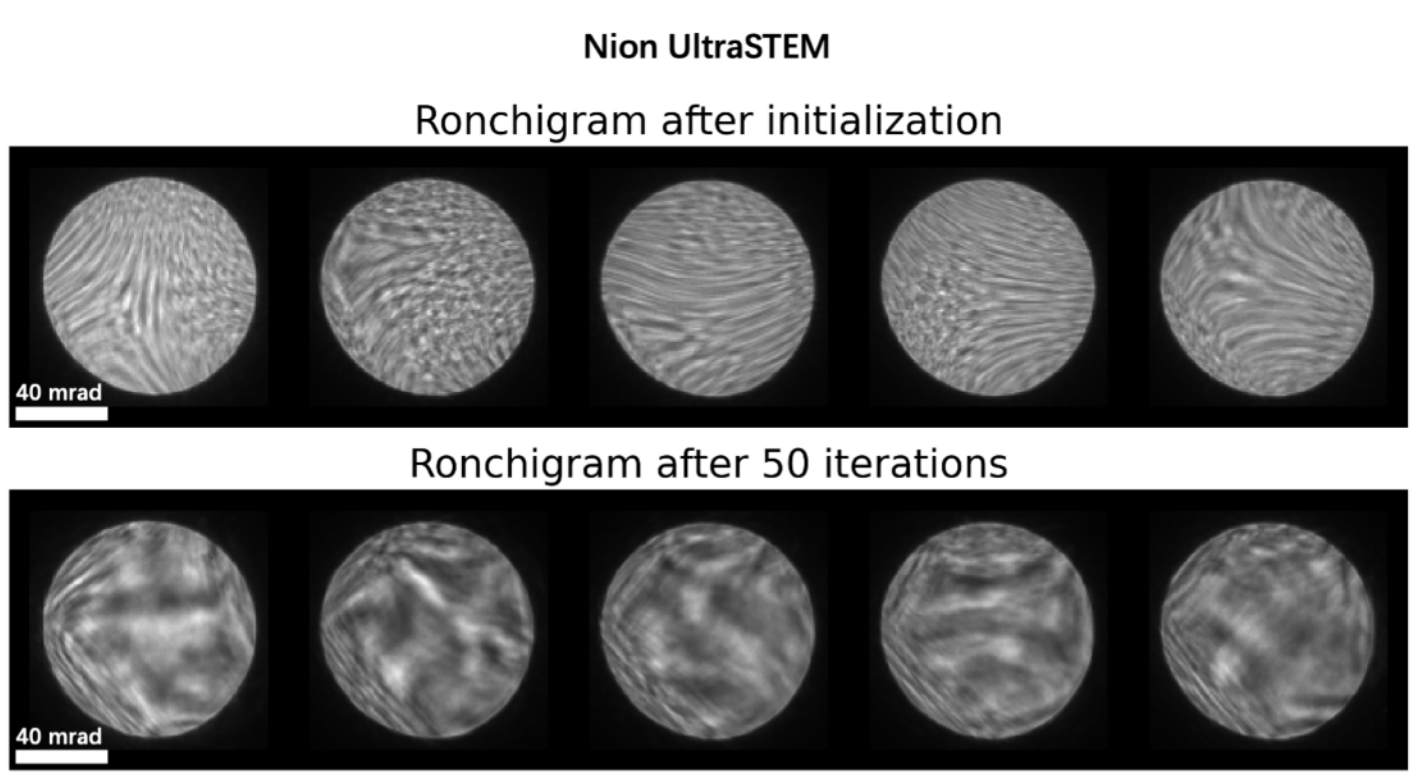

FIG. A1: Ronchigram appearance before and after online Bayesian optimization of Nion UltraSTEM. The Nion microscope used a larger aperture size of 45 mrad which would include more higher order aberrations in the Ronchigram, but also had some intrinsic limitations in the lens design that prohibited perfect tuning even at smaller angles.

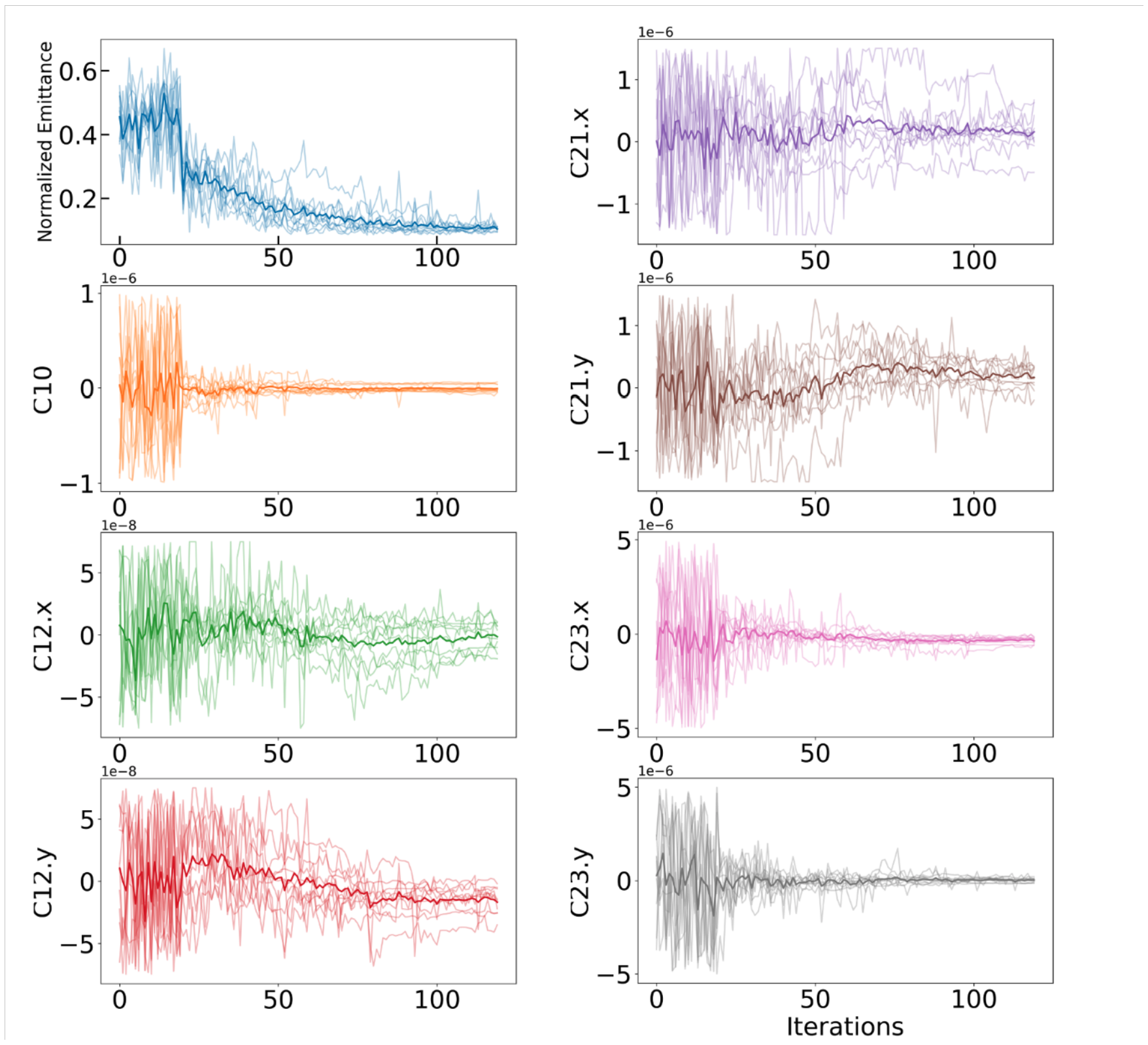

FIG. A2: Iterations of different aberration tuning commands for the Nion UltraSTEM. Most coefficient values stabilize after roughly 20 iterations.

We first inspect the popular RBF kernel.

$$k_{\mathrm{RBF}}(\mathbf{x_1},\mathbf{x_2}) = \exp\left(-\frac{1}{2}(\mathbf{x_1}-\mathbf{x_2})^{\top}\Theta^{-2}(\mathbf{x_1}-\mathbf{x_2})\right) \tag{B1}$$

where $\mathbf{x}_1$ and $\mathbf{x}_2$ are a pair of inputs, here current values of the hextuples. $\Theta$ is a lengthscale parameter. The RBF kernel has values in $[0,1)$, which already has the interpretation of a similarity measure. The exponential essentially maps the inputs to an infinite dimensional space, making it more flexible than linear or polynomial kernels. Though simple, we see being infinitely differentiable the RBF kernel is enforcing a strong assumption of smoothness, which in many cases is not warranted. When discontinuity exists, the RBF kernel would over-smooth the region because it is assuming a gradual decay in similarity between the inputs over a distance. A solution to this is the Matern kernels.

$$k_{\mathrm{Matern}}(\mathbf{x_1},\mathbf{x_2}) = \frac{2^{1-v}}{\Gamma(v)}(\sqrt{2v}d)^{v}K_v(\sqrt{2v}d) \tag{B2}$$

where $d = (\mathbf{x_1} - \mathbf{x_2})^{\top}\,\Theta^{-2}\,(\mathbf{x_1} - \mathbf{x_2})$ is the distance between $x_1$ and $x_2$ scaled by the length scale parameter $\Theta$. $K_v$ is a modified Bessel function. $v$ is a smoothness parameter which effectively controls the level of smoothness. As $\mu \to \infty$ we recover exactly the RBF kernel. When $v$ is half-integer, the Matern kernel has a nice mathematical form as the product of an exponential function and a polynomial function of order d.

For added flexibility, we also propose to use spectral mixture base kernels (Wilson and Adams, 2013): The spectral mixture (SM) kernel, which forms an expressive basis for all stationary covariance functions, can discover quasi-periodic stationary structure with an interpretable and succinct representation, while the deep learning transformation $g(x, w)$ captures non-stationary and hierarchical structure,

$$k_{\mathrm{SM}}(\mathbf{x}_1, \mathbf{x_2} \mid \boldsymbol{\theta}) = \sum_{q=1}^{Q} a_q \frac{\left|\Sigma_q\right|^{\frac{1}{2}}}{(2\pi)^{\frac{D}{2}}} \exp\left(-\frac{1}{2} \parallel \Sigma_q^{\frac{1}{2}}(\mathbf{x}_1 - \mathbf{x_2})\right|^2\right) \cos\langle \boldsymbol{x_1} - \boldsymbol{x_2}, 2\pi\boldsymbol{\mu}_q\rangle \qquad \text{(B2)}$$

The parameters of the spectral mixture kernel $\theta = \{a_q, \Sigma_q, \mu_q\}$ are mixture weights, band-widths (inverse length-scales), and frequencies.

**Appendix C: Additional benchmark on choices of BO hyperparameters**

We supplement the benchmarking on GPT-6D simulation with additional choices of kernels, acquisition functions and hyperparameters. We show all deep kernel implementations outperform the isotropic kernels. Matern kernel outperforms RBF kernel at an earlier stage, but eventually converge to the same level as a result of both being isotropic as shown in Figure C1.

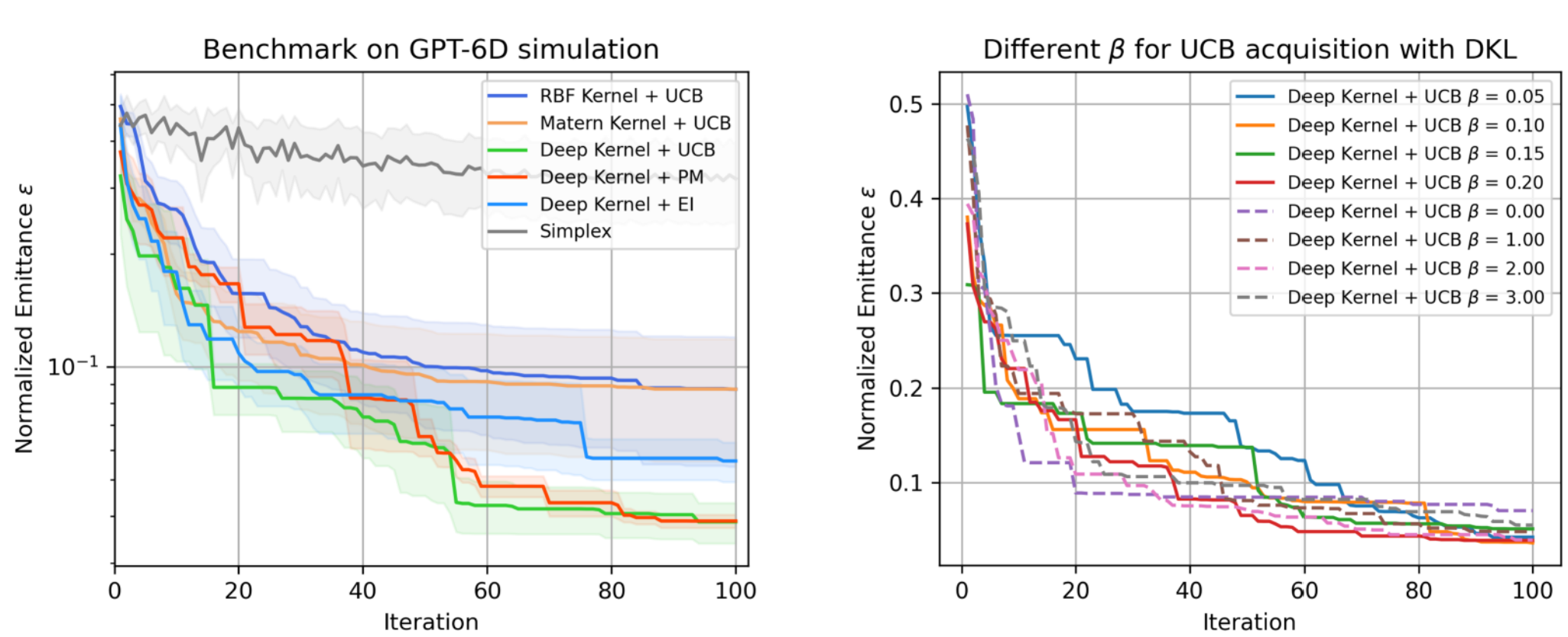

FIG. C1: Additional benchmark results of the GPT-6D simulation. Left: Performance of different kernels and acquisition functions averaged over 10 repetitions. Right: DKBO with varying $\beta$ for UCB acquisition function.